\documentclass{Interspeech2024}

\interspeechcameraready

\usepackage{amsmath,graphicx}
\usepackage[table]{xcolor}
\usepackage{booktabs}
\usepackage{multirow}
\usepackage{multicol}
\usepackage{amssymb}
\usepackage[usestackEOL]{stackengine}
\usepackage[backend=bibtex,citestyle=numeric-comp,bibstyle=ieee,defernumbers=true,giveninits=true,doi=false,isbn=false,url=false,eprint=false,minbibnames=2,maxbibnames=2]{biblatex}
\AtEveryBibitem{
\clearname{editor}
\clearfield{pages}
\clearfield{volume}
\clearfield{number}
}

\usepackage{pifont}
\newcommand{\cmark}{\ding{51}}%
\newcommand{\xmark}{\ding{55}}%
\def\ie{\emph{i.e.}}

\def\eg{\emph{e.g.}}

\makeatletter
\g@addto@macro{\endtabular}{\rowfont{}}
\makeatother
\newcommand{\rowfonttype}{}
\newcommand{\rowfont}[1]{
\gdef\rowfonttype{#1}#1\ignorespaces%
}
\makeatother

\newcommand\blfootnote[1]{%
  \begingroup
  \renewcommand\thefootnote{}\footnote{#1}%
  \addtocounter{footnote}{-1}%
  \endgroup
}

\title{RT-LA-VocE: Real-Time Low-SNR Audio-Visual Speech Enhancement}

\name[affiliation={1}]{Honglie}{Chen$^{*}$}
\name[affiliation={2}]{Rodrigo}{Mira$^{*}$}
\name[affiliation={1,2}]{Stavros}{Petridis}
\name[affiliation={1,2}]{Maja}{Pantic}

\address{
  $^1$Meta AI, UK \quad 
  $^2$Imperial College London, UK
  }
\email{hongliechen@meta.com, \{rs2517, stavros.petridis04, m.pantic\}@imperial.ac.uk}

\begin{document}
\ninept
\maketitle

\begin{abstract}
In this paper, we aim to generate clean speech frame by frame from a live video stream and a noisy audio stream without relying on future inputs.
To this end, we propose RT-LA-VocE, which completely re-designs every component of LA-VocE, a state-of-the-art non-causal audio-visual speech enhancement model, to perform causal real-time inference with a 40\,ms input frame. 
We do so by devising new visual and audio encoders that rely solely on past frames, replacing the Transformer encoder with the Emformer, and designing a new causal neural vocoder \emph{C-HiFi-GAN}. 
On the popular AVSpeech dataset, we show that our algorithm achieves state-of-the-art results in all real-time scenarios.
More importantly, each component is carefully tuned to minimize the algorithm latency to the theoretical minimum (40\,ms) while maintaining a low end-to-end processing latency of 28.15\,ms per frame, enabling real-time frame-by-frame enhancement with minimal delay.

\end{abstract}
\blfootnote{$^*$Equal contribution.}

\vspace{-0.1cm}
\section{Introduction}
\vspace{-0.1cm}
\label{sec:intro}
When conversing in the real world, our speech is often mixed with other undesirable signals, such as noise from cars on the road, or another conversation happening nearby. 
For this reason, the concept of noise reduction in real time has long been considered an attractive prospect for modern speech transmission frameworks~\cite{benesty2006speech}. 
With this in mind, several deep learning-based methods have recently been proposed to tackle this challenge by adapting existing non-causal models~\cite{DBLP:conf/interspeech/DefossezSA20} or designing new architectures tailored specifically for real-time enhancement~\cite{DBLP:conf/icassp/0004W19,DBLP:conf/interspeech/ThakkerEYW22}. 
These models achieve impressive noise reduction performance in high-SNR (signal-to-noise ratio) environments but do not experiment with substantially noisier scenarios, where the performance of audio-only models often sharply deteriorates~\cite{DBLP:conf/ica/GaoDXLDL15,DBLP:conf/mlsp/BirnieSAG21,DBLP:conf/interspeech/HaoSW0B19}. 
Furthermore, these single-channel audio-only models are unable to remove interfering speech in a multi-talker environment, limiting their potential applications.

\begin{figure}[t]
\centering 
\includegraphics[width=\linewidth]{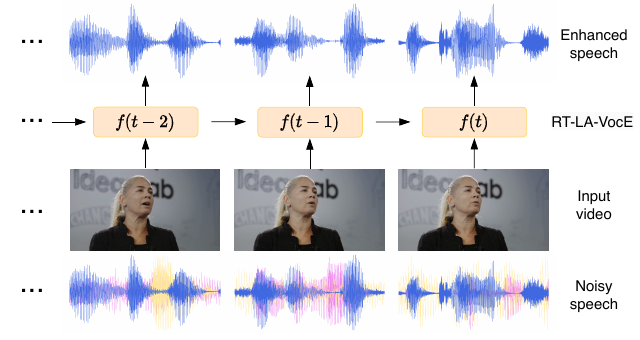}
\vspace{-0.8cm}
\caption{RT-LA-VocE's real-time audio-visual speech enhancement approach. The model processes 40\,ms frames in real time.}
\vspace{-0.6cm}
\label{overview_fig}
\end{figure}

In recent years, the increased bandwidth of modern devices has paved the way for video streaming as a ubiquitous form of communication, particularly in professional environments~\cite{doi:10.1177/10464964211015286}. Compared to audio-only pipelines, video streaming adds a visual stream of the speaker's face, which can contain valuable verbal information, as demonstrated by various studies in audio-visual learning~\cite{DBLP:conf/icassp/PetridisSMCTP18,DBLP:conf/iclr/Haliassos0MPP23}. 
In light of this, emerging approaches have borrowed from speech enhancement and lipreading literature to develop real-time audio-visual speech enhancement (AVSE) frameworks that effectively outperform their audio-only counterparts~\cite{DBLP:conf/icassp/PanTX021,DBLP:conf/cvpr/GaoG21,DBLP:conf/cvpr/YangMKAR22,DBLP:journals/corr/abs-2212-11377}. 
Most works focus on real-time inference on edge devices equipped with low-end CPUs, such as \cite{DBLP:journals/inffus/GogateDAH20,DBLP:journals/corr/abs-2112-09060,gogate22_spsc}, which present a real-time AVSE model trained on studio-recorded datasets (GRID~\cite{grid} and TCD-TIMIT~\cite{7050271}), and \cite{10094724}, which extends an existing real-time speech enhancement model~\cite{DBLP:conf/interspeech/ThakkerEYW22} by adding a lightweight video encoder and a multi-stage fusion module. Alternatively, \cite{DBLP:conf/eccv/MontesinosKH22} proposes a transformed-based model running on a high-end GPU, leveraging the target speaker's facial landmarks. 
These real-time audio-visual approaches succeed in outperforming their audio-only counterparts, but fail to experiment with low-SNR scenarios featuring multiple interfering speakers and do not directly attempt to compete with non-causal AVSE models~\cite{DBLP:conf/icassp/PanTX021,DBLP:conf/cvpr/GaoG21,DBLP:conf/cvpr/YangMKAR22,DBLP:journals/corr/abs-2212-11377,lavoce}. 
More importantly, existing AVSE models are limited in terms of deployment as they benchmark inference speed on long sequences, \eg, 10s in \cite{DBLP:conf/eccv/MontesinosKH22} and 3s in \cite{10094724}, and do not report the model's processing latency per frame. In fact, frame-by-frame enhancement, which is necessary for online enhancement in real-world scenarios, is almost entirely neglected by existing real-time AVSE works in both their methodology and experimental procedure. 

To bridge the aforementioned gaps, we aim to explore real-time low-SNR audio-visual speech enhancement with minimal frame-by-frame latency (shown in Figure~\ref{overview_fig}), enabling live noise reduction in real-word video streaming systems. 
To this end, we develop a causal AVSE model based on the state-of-the-art non-causal model LA-VocE~\cite{lavoce}, which consists of a large Transformer-based spectrogram enhancer and a neural vocoder (HiFi-GAN V1~\cite{DBLP:conf/nips/KongKB20}). 
We do so by building new, fully causal video and audio encoders, replacing the transformer with the recently proposed Emformer~\cite{DBLP:conf/icassp/ShiWWYC0LS21} and proposing a new vocoder -- \emph{C-HiFi-GAN} -- which can synthesize raw waveforms from spectrograms without depending on future information.

Concretely, we make the following contributions: 
(i) We introduce a novel causal AVSE architecture designed for low-SNR conditions - RT(Real-Time)-LA-VocE; 
(ii) we outperform all other causal approaches and perform competitively with the state-of-the-art non-causal models; more importantly 
(iii), we reduce the algorithm latency to the minimum for real-time AVSE models, \ie, 1 video frame (40\,ms), so a negligible delay is introduced during real-time inference; and finally 
(iv), we achieve a total processing latency of 28.15\,ms per frame, considerably less than the time needed to obtain the next frame (40\,ms) on a server-side GPU, demonstrating the model's capability for real-world streaming applications.

\vspace{-0.1cm}
\section{Methodology} \label{sec:method}
\begin{figure*}[t]
\centering 
\includegraphics[width=\linewidth]{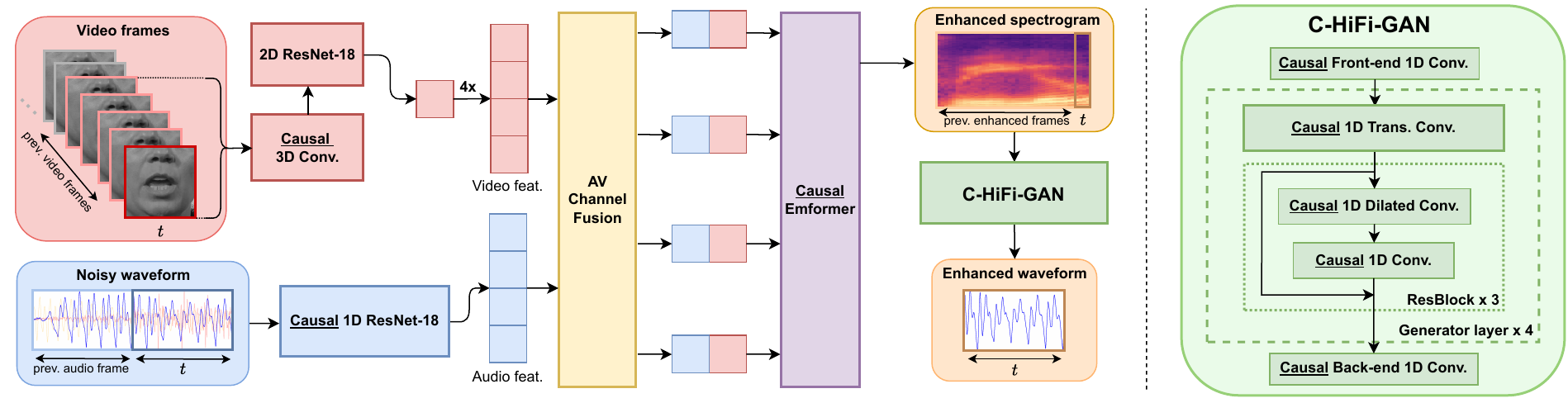}
\vspace{-0.6cm}
\caption{Detailed overview of RT-LA-VocE's inference pipeline for each time-step $t$ (40 ms). RT-LA-VocE receives five video frames, which are passed through our ResNet-based visual encoder, and two raw audio frames, which are encoded via our causal 1D ResNet-18. The resulting features are concatenated channel-wise and fed into the Emformer, which models the temporal dynamics with previous time-steps. This is followed by a linear layer that predicts the four enhanced spectrogram frames. Finally, these frames are combined with past predictions and fed into \emph{C-HiFi-GAN}, which generates the corresponding waveform. }
\label{method_fig}
\vspace{-0.6cm}
\end{figure*}
\vspace{-0.1cm}
\subsection{LA-VocE}
\vspace{-0.1cm}
We start by describing the multi-stage approach introduced in LA-VocE which consists of an audio-visual spectrogram enhancer and an adapted version of HiFi-GAN V1~\cite{DBLP:conf/nips/KongKB20}. 
In the first stage, given a video clip and a spectrogram, the visual and auditory representations are computed using a 2D ResNet-18 preceded by a 3D front-end layer (as in \cite{DBLP:conf/icassp/PetridisSMCTP18}) and a linear encoder, respectively. 
To jointly model both modalities, the two sets of features are concatenated channel-wise and passed through a 12-block Transformer encoder~\cite{NIPS2017_3f5ee243}. 
The output from the Transformer is then fed to a linear projection layer to generate the enhanced spectrogram. This model is trained by minimising the L1 loss between the enhanced and clean spectrograms.
In the second stage, a neural vocoder is leveraged to translate the enhanced spectrogram into a waveform - HiFi-GAN V1~\cite{DBLP:conf/nips/KongKB20}. 
In short, HiFi-GAN's generator is a fully convolutional neural network consisting of 4 blocks of transposed convolutional layers and multi-receptive field fusion modules (MRF).
It is trained using a combination of comparative and adversarial losses, which leverage an essemble of multi-period and multi-scale discriminators.

\par{\noindent \bf Limitations.} 
The goal of this paper is to enable real-time speech enhancement with minimal latency, \ie, to produce a causal model that can sequentially generate enhanced audio frame by frame from a live audio-visual stream without relying on future inputs. 
To achieve this, the most straightforward solution is to na\"ively send frame-by-frame inputs to the original LA-VocE. 
However, this model is designed to rely heavily on future information along several modules, namely: the 3D convolution in the visual encoder, which leverages 2 future video frames; 
the STFT (short-time Fourier transform) used to compute the mel-spectrogram, whose window extends 15~ms into the future; 
the multi-head attention layers in the spectrogram enhancer, where the output for time-step $t$ depends on all past and future time-steps; 
and the HiFi-GAN, whose convolutions gradually bleed future information into their outputs. 
Therefore, this na\"ive approach leads to severly deteriorated performance, as we will demonstrate in Table~\ref{tab:on_off_inference}.

\vspace{-0.1cm}
\subsection{Real-time LA-VocE} 
\label{sec:method_rt_lavoce}
\vspace{-0.1cm}
To resolve these challenges, the model should be constrained to depend exclusively on past information. 
In this section, we describe our key changes to each module of the original LA-VocE and show our proposed RT-LA-VocE is completely causal with minimal algorithm latency. 
In Figure~\ref{method_fig}, we illustrate our end-to-end real-time inference pipeline in detail. 

\par{\noindent \bf Video encoder.} 
As proposed in \cite{DBLP:conf/ssw/OordDZSVGKSK16}, a convolutional layer can be made causal by adding $p_{conv}$ padding frames to the left of the input and removing an equivalent amount of frames from the right of the output (to maintain the same output dimension):
\begin{align}
    p_{conv}=\lfloor \frac{k}{2} \rfloor \cdot d,
\end{align}
where $k$ and $d$ denote the kernel size and dilation of the convolution, respectively. As shown in Figure~\ref{method_fig}, we conduct this change on the $3$D convolution in the video encoder, which creates a causal model that depends only on past video frames. This change reduces the algorithm latency of the video encoder from 120\,ms (3 frames) to 40\,ms (1 frame).

\par{\noindent \bf Audio encoder.} 
LA-VocE encodes audio by converting the raw waveform into a mel-spectrogram, and applying a linear layer on the resulting magnitudes. The STFT operation used to obtain the mel-spectrogram introduces extra algorithm latency since the input raw waveform is padded on both sides by $p_{stft}$:
\begin{align}
    p_{stft}=\frac{w - h}{2},
\end{align}
where $w$ specifcies the window size (40~ms) and $h$ specifies the hop size (10~ms), meaning that the STFT window depends on 15~ms of future information. There are two potential solutions to remove this extra latency
: i) pad the left side of the waveform by $2p$ (analogous to the convolution trick mentioned above); or ii) apply a causal encoder on the raw waveform instead. We adapt the 1D ResNet-18 from \cite{DBLP:conf/iclr/Haliassos0MPP23} using the causal padding trick on all convolutions to form an entirely causal raw audio encoder. We also increase the temporal resolution from 25\,Hz (one audio feature per video frame) to 100\,Hz (four audio features per video frame) by adjusting the kernel size and stride in the final pooling layer. We compare these causal encoder variants in Table~\ref{tab:ablation}.

\par{\noindent \bf Emformer.} 
In addition to the convolutional layers mentioned above, the Transformer encoder also explicitly takes into account future information via the multi-head attention operation.
In order to achieve a causal temporal encoder, we replace the Transformer encoder with the recently proposed Emformer~\cite{DBLP:conf/icassp/ShiWWYC0LS21}.
It adapts the original Transformer architecture for causal inference by computing attention over the current frame and a sequence of previous frames known as the left context, removing the attention layers' reliance on future time-steps.

\par{\noindent \bf C-HiFi-GAN.} 
Similar to the 3D layer in the video encoder, the convolutions and transposed convolutions in HiFi-GAN also leak future information into the temporal representations, making the model non-causal. 
We design \emph{C-HiFi-GAN} (Causal HiFi-GAN) by employing the causal padding trick for all convolutions in each MRF module of the original HiFi-GAN V1, as well as the four transposed convolutions by padding with:
\begin{align}
    p_{conv_{trans}}=\lfloor \frac{s}{2} \rfloor + s \bmod 2,
\end{align}
where $s$ denotes the stride of the transposed convolution.


\vspace{-0.1cm}
\section{Experimental Setup}
\vspace{-0.1cm}
\subsection{Datasets and evaluation metrics}
\label{sec:datasets}
\vspace{-0.1cm}
Following the original LA-VocE~\cite{lavoce}, we sample clean and interfering speech from AVSpeech~\cite{Ephrat2018LookingTL} and noise from the DNS challenge noise dataset~\cite{reddy2020interspeech}. 
For consistency, we use the same train/test split as the original LA-VocE.
The level of speech interference (\ie, audio from other speakers in the background) and background noise are controlled by the signal-to-interference ratio (SIR) and signal-to-noise ratio (SNR), respectively:
\begin{align}
    \text{SIR} = \frac{P_{signal}}{P_{interference}}, & \qquad \qquad \quad \text{SNR} =  \frac{P_{signal}}{P_{noise}},
\end{align}
where $P_{x}$ refers to the power of the raw signal $x$. 
We follow the approach in \cite{lavoce} to crop the 96$\times$96-sized mouth region from each video\footnote{Only non-Meta authors conducted any of the dataset preprocessing (no dataset pre-processing took place on Meta’s
servers or facilities).}.
Additionally, a low-latency causal mouth cropping pipeline is created using MediaPipe~\cite{DBLP:journals/corr/abs-1906-08172}. 
We compare these two methods for real-time inference in Table~\ref{tab:lightweight_crop}.
We measure speech quality using MCD~\cite{407206}, PESQ-WB~\cite{DBLP:conf/icassp/RixBHH01}, and ViSQOL~\cite{DBLP:conf/qomex/ChinenLSGOH20}, and speech intelligibility using STOI~\cite{DBLP:journals/taslp/TaalHHJ11} and its extended version ESTOI~\cite{DBLP:journals/taslp/JensenT16}. 
Unlike \cite{lavoce}, we present the raw metrics computed on the enhanced samples rather than the improvement over the noisy baseline to clearly illustrate their similarity with the clean speech in absolute terms. 
We evaluate all models under the same three noise conditions 1, 2, and 3 proposed in \cite{lavoce}, featuring 1, 3, and 5 background noises at 0, -5, and -10 dB, and 1, 2, and 3 interfering speakers at 0, -5, and -10 dB SIR, respectively.
\vspace{-0.1cm}
\subsection{Comparison models}
\vspace{-0.1cm}
We compare our work with three non-causal AVSE models: LA-VocE~\cite{lavoce}, MuSE~\cite{DBLP:conf/icassp/PanTX021}, and VisualVoice~\cite{DBLP:conf/cvpr/GaoG21}. 
In addition, we adapt two speech enhancement models 
-- GCRN~\cite{DBLP:journals/taslp/TanW20} and Demucs~\cite{DBLP:journals/corr/abs-1911-13254} -- for real-time audio-visual enhancement by adding a causal ResNet-based encoder to each, as in our model, and concatenating the resulting visual features with the audio features in the bottleneck, as in \cite{lavoce}. 
We use the causal version of Demucs presented in \cite{DBLP:conf/interspeech/DefossezSA20}.
We also compare with two real-time audio-only models: GCRN and an audio-only version of RT-LA-VocE, where we remove the visual stream entirely. 
All models are trained on the dataset presented in Section~\ref{sec:datasets} with optimization parameters based on the ones used for our spectrogram enhancer (described in Section~\ref{sec:train_details}). 
As in \cite{lavoce}, MuSE is trained using the loss ensemble proposed in \cite{DBLP:journals/corr/abs-1911-13254}.
\vspace{-0.4cm}
\subsection{Implementation details}
\begin{table*}[t]
\centering
\setlength{\tabcolsep}{2pt}
\resizebox{\linewidth}{!}{%
\begin{tabular}{@{}lcccccccccccccccccc@{}}
\toprule
\multicolumn{1}{l}{\multirow{3}{*}{\vspace{0.2cm}Model}} & \multirow{3}{*}{\vspace{0.2cm}Input} & \multirow{3}{*}{\vspace{0.2cm}Online} & \multirow{3}{*}{\vspace{0.2cm}\Centerstack{Proc. latency \\ (ms)}} & \multicolumn{5}{c}{Noise condition 1} & \multicolumn{5}{c}{Noise condition 2} & \multicolumn{5}{c}{Noise condition 3} \\
\cmidrule(lr){5-9} \cmidrule(lr){10-14} \cmidrule(lr){15-19}
\multicolumn{4}{c}{} & \multicolumn{1}{c}{ MCD$\downarrow$} & \multicolumn{1}{c}{ PESQ$\uparrow$}& \multicolumn{1}{c}{ ViSQOL$\uparrow$}& \multicolumn{1}{c}{ STOI$\uparrow$}& \multicolumn{1}{c}{ ESTOI$\uparrow$}& \multicolumn{1}{c}{ MCD$\downarrow$} & \multicolumn{1}{c}{ PESQ$\uparrow$}& \multicolumn{1}{c}{ ViSQOL$\uparrow$}& \multicolumn{1}{c}{ STOI$\uparrow$}& \multicolumn{1}{c}{ ESTOI$\uparrow$}& \multicolumn{1}{c}{ MCD$\downarrow$} & \multicolumn{1}{c}{ PESQ$\uparrow$}& \multicolumn{1}{c}{ ViSQOL$\uparrow$}& \multicolumn{1}{c}{ STOI$\uparrow$}& \multicolumn{1}{c}{ ESTOI$\uparrow$} \\ \midrule
 Baseline & - & - & - &  10.81&1.088&1.168&0.545&0.367&12.05&1.105&1.053&0.386&0.195& 12.47&1.160&1.035&0.281&0.093 \\ \midrule
AV-GCRN&AV&\xmark& - & 9.297&1.514&1.710&0.773&0.614&10.457&1.215&1.483&0.630&0.424&10.989&1.112&1.271&0.462&0.245\\
AV-Demucs&AV& \xmark& -&5.260&1.818&1.852&0.813&0.661&6.522&1.383&1.484&0.692&0.495&7.608&1.172&1.334&0.543&0.323\\
MuSE&AV& \xmark& - &5.282&	1.875&1.847&0.821&0.666 &6.736&1.402&1.462&0.694&0.484 &8.285&1.171&1.277&0.512&0.275 \\
VisualVoice&AV&\xmark& -&6.869&1.689&1.815&0.794&0.637&8.515&1.272&1.419&0.637&0.431&9.806&1.114&1.283&0.460&0.252\\
LA-VocE &AV&\xmark& -&\underline{4.177}&\underline{2.017}&\underline{2.265}&\underline{0.839}&\underline{0.700}& \underline{5.217}&\underline{1.621}&\underline{1.743}&\underline{0.765}&\underline{0.592}&\underline{6.320}&\underline{1.317}&\underline{1.480}&\underline{0.651}&\underline{0.451}\\ 
\midrule 
RT-GCRN &A&\cmark& 4.12$\pm$0.10&11.275&1.129&1.260&0.495&0.327&11.682&1.102&1.212&0.362&0.176&12.066&1.151&1.238&0.257&0.086 \\
RT-LA-VocE & A & \cmark& 17.80$\pm$0.03 & 8.761 & 1.148 & 1.205 & 0.490 & 0.315 & 9.934 & 1.112 & 1.115 & 0.376 & 0.172 & 10.585 & 1.166 & 1.119 & 0.293 & 0.086 \\ 
RT-AV-GCRN&AV &\cmark& 6.56$\pm$0.15&9.713&1.496&1.691&0.769&0.607&10.675&1.209&1.472&0.622&0.416&11.087&1.117&1.262&0.452&0.237\\
RT-AV-Demucs&AV & \cmark& 4.87$\pm$0.13 & 6.228&1.408 &1.464 &0.729 &0.550 & 7.363&1.202 &1.281& 0.591&0.375 &8.361 &1.108 &1.239 &0.443 &0.222 \\
RT-LA-VocE & AV & \cmark& 20.88$\pm$0.04 & \textbf{4.653} & \textbf{1.741} & \textbf{2.050} & \textbf{0.800} & \textbf{0.649} & \textbf{5.737} & \textbf{1.402} & \textbf{1.615} & \textbf{0.701} & \textbf{0.516} & \textbf{6.799} & \textbf{1.199} & \textbf{1.402} & \textbf{0.568} & \textbf{0.365} \\ 
\bottomrule
\end{tabular}
}
\caption{Comparison between RT-LA-VocE and other speech enhancement methods for different noise conditions. The best results for offline inference (among the non-causal models) are \underline{underlined}, and the best online results (among the causal models) are highlighted in \textbf{bold}. ``Proc. latency'' denotes the processing latency per frame (40\,ms) during inference. \label{tab:comparison}}
\vspace{-0.6cm}
\end{table*}
\begin{table*}[!htb]
  \resizebox{1\linewidth}{!}{
  \begin{tabular}{@{}llccccccccc@{}}
    \toprule
    Audio encoder & Temporal model & Spec. inv. method  & MCD $\downarrow$ & PESQ $\uparrow$ &  ViSQOL $\uparrow$ & STOI $\uparrow$ & ESTOI $\uparrow$ & Alg. latency (ms) &  Proc. latency (ms)  & \# Params (M)\\
    \midrule 
    Spec. + lin. layer& Emformer ($l_c=32$)& Noisy phase & 6.342&1.278&1.457&0.589&0.39 & 55 (40 + 15) & 13.65$\pm$0.13 & 96.6 \\
    Mel-spec. + lin. layer & Emformer ($l_c=32$) & Noisy phase & 5.962	& 1.33 & 1.54 & 0.651 & 0.448 & 55 (40 + 15) & 19.07$\pm$0.25 & 96.3 \\
    Mel-spec. + lin. layer & Emformer ($l_c=32$) & Griffin-Lim & 6.468&1.236&1.39&0.636&0.446 & 55 (40 + 15) & 30.86$\pm$0.31 & 96.3 \\
    Mel-spec. + lin. layer & Emformer ($l_c=32$) & C-HiFi-GAN &5.906&1.391&1.567&0.694&0.504 & 55 (40 + 15) & 18.81$\pm$0.04 & 110 \\
    \midrule
    Causal mel-spec.+ lin. layer & Emformer ($l_c=32$) & C-HiFi-GAN & 6.145 & 1.302 & 1.463 & 0.645 & 0.456 & \textbf{40} & 19.44$\pm$0.04 & 110 \\
    Causal 1D ResNet (25\,Hz) & Emformer ($l_c=32$) & C-HiFi-GAN & 5.988 & 1.261 & 1.421 & 0.614 & 0.436 & \textbf{40} & 19.86$\pm$0.04 & 114 \\
    Causal 1D ResNet (100\,Hz) & Emformer ($l_c=32$) & C-HiFi-GAN & 5.752&1.397&1.592&0.696&0.510 & \textbf{40} & 20.01$\pm$0.04 & 114 \\
    Causal 1D ResNet (100\,Hz) & Emformer ($l_c=64$) & C-HiFi-GAN & \textbf{5.737}&\textbf{1.402}&\textbf{1.615}&\textbf{0.701}&\textbf{0.516} & \textbf{40} & 20.88$\pm$0.04 & 114 \\
    \bottomrule
  \end{tabular}
  }
  \caption{Ablation on the main architectural components of RT-LA-VocE (noise condition 2). $l_c$ denotes left context length. \label{tab:ablation}}
  \vspace{-0.9cm}
\end{table*}
\begin{table}[t]
  \centering
  \setlength{\tabcolsep}{2pt}
  \resizebox{1\linewidth}{!}{
  \begin{tabular}{@{}cccccccc@{}}
    \toprule
    \multicolumn{2}{c}{Backbone} & \multirow{2}{*}{\vspace{-0.15cm}Online} &  \multirow{2}{*}{\vspace{-0.15cm}MCD $\downarrow$} & \multirow{2}{*}{\vspace{-0.15cm}PESQ $\uparrow$} &  \multirow{2}{*}{\vspace{-0.15cm}ViSQOL $\uparrow$} & \multirow{2}{*}{\vspace{-0.15cm}STOI $\uparrow$} & \multirow{2}{*}{\vspace{-0.15cm}ESTOI $\uparrow$}\\
    \cmidrule(lr){1-2}
    Temp. & Spec. inv. & \multicolumn{6}{c}{} \\
    \midrule 
    Transformer & HiFi-GAN & \xmark  & 5.248&1.64&1.826&0.77&0.606   \\
    Transformer & C-HiFi-GAN & \xmark  &5.285&1.633&1.815&0.773&0.602    \\
    Emformer & HiFi-GAN & \xmark   & 5.696 & 1.422 & 1.628 & 0.710 & 0.527  \\
    Emformer & C-HiFi-GAN & \xmark   & 5.737 & 1.402 & 1.615 & 0.701 & 0.516 \\ \midrule
    Transformer & HiFi-GAN & $\checkmark$ & 12.394&1.073&1.052&0.125&0.036  \\
    Transformer & C-HiFi-GAN & $\checkmark$ & 12.502&1.09&1.035&0.12&	0.034  \\
    Emformer & HiFi-GAN  & $\checkmark$ & 6.027 & 1.299 & 1.445 & 0.687 & 0.502  \\
    Emformer & C-HiFi-GAN  & $\checkmark$ & \textbf{5.737} & \textbf{1.402} & \textbf{1.615} & \textbf{0.701} & \textbf{0.516}  \\
    \bottomrule
  \end{tabular}
  }
  \caption{Online and offline inference results with causal and non-causal models (noise condition 2).}
  \vspace{-0.7cm}
\label{tab:on_off_inference}
\end{table}
\begin{table}[t]
\centering
\footnotesize
\setlength{\tabcolsep}{4pt}
\resizebox{\linewidth}{!}{
\begin{tabular}{lcccccc}
\toprule
Mouth crop. method & Latency (ms) & MCD$\downarrow$ & PESQ $\uparrow$ &  ViSQOL$\downarrow$ & STOI $\uparrow$ &  ESTOI$\downarrow$ \\ \midrule
LA-VocE~\cite{lavoce} & 92.65$\pm$2.56 & \textbf{5.737} & \textbf{1.402} & \textbf{1.615} & \textbf{0.701} & \textbf{0.516}  \\
MediaPipe~\cite{DBLP:journals/corr/abs-1906-08172} & \textbf{7.27$\pm$0.07} & 5.910 & 1.380 & 1.577 & 0.689 & 0.503 \\ \bottomrule
\end{tabular}}%
{\vspace{0pt}
\caption{Quantitative results on different mouth crops (noise condition 2). LA-VocE's crops are smoothed using future frames, while MediaPipe's crops are completely causal.}\label{tab:lightweight_crop}}
\vspace{-1.0cm} 
\end{table}
\label{sec:train_details}
\vspace{-0.1cm}
\par{\noindent \bf Architectural Details.} 
We use window size 640, hop size 160, frequency bin size 640, and 80 mel bands to compute the mel-spectrograms.
Our Emformer consists of 12 blocks featuring 12 attention heads, a hidden unit dimension of 768, a segment length of 4, a left context length of 64 and a memory bank length of 0. Note, in our experiments, no noticeable improvement was observed with memory bank lengths $>$~0.
We use 4 blocks of transposed convolutional layers in \emph{C-HiFi-GAN} with upsampling factors of $8\times$, $5\times$, $2\times$, and $2\times$.
MRF is made of 3 residual blocks featuring dilated and non-dilated convolutional layers with kernel sizes $3$, $7$, and $11$. During testing, each input audio frame's duration is 40\,ms, \ie, a single video frame.

\par{\noindent \bf Training curriculum.} 
To train the spectrogram enhancer, we use AdamW with learning rate $7\times10^{-4}$, $\beta_1 = 0.9$, $\beta_2 = 0.98$, and weight decay $3\times10^{-2}$. 
We train for 150 epochs using linear learning rate warmup for the first 10\,\% of training, and applying a cosine decay schedule for the remaining epochs. 
Separately, we train \emph{C-HiFi-GAN} using AdamW with learning rate $2\times10^{-4}$, $\beta_1 = 0.8$, and $\beta_2 = 0.99$.
We train for a total of $1$M steps while decaying the learning rate by a factor of $0.999$ after each epoch. 
During training, we apply standard data augmentations such as random cropping, horizontal flipping, erasing, and time masking, as in \cite{lavoce}. The latencies in the following section are measured on an NVIDIA RTX 2080 Ti GPU with an Intel Core i7-9700K CPU, and averaged over 1000 inference steps.
\vspace{-0.1cm}
\section{Results}
\vspace{-0.1cm}
\subsection{Comparison with the state-of-the-art}
\vspace{-0.1cm}
We compare our results with recent causal and non-causal speech enhancement models across three challenging noise conditions in Table~\ref{tab:comparison}. Overall, it is clear that RT-LA-VocE yields substantial improvements compared to the noisy baseline, halving the MCD and achieving relative improvements greater than 50\,\% for STOI and ESTOI throughout all 3 noise conditions.  Conversely, RT-AV-GCRN and RT-AV-Demucs are only able to achieve considerable improvements for noise condition 1 and feature a sharp decline in noise conditions 2 and 3. 
While RT-LA-VocE does not exactly match LA-VocE's performance, it is remarkably competitive with this state-of-the-art model on both quality and intelligibility without leveraging any future information. Furthermore, RT-LA-VocE outperforms two recent non-causal models (AV-GCRN and VisualVoice) across all noise conditions, and is competitive with AV-Demucs and MuSE, achieving better performance on noise conditions 2 and 3. Finally, we highlight that both audio-only speech enhancement models fail to achieve noticeable improvements in quality or intelligibility. Indeed, as mentioned in Section~\ref{sec:intro}, with only a single stream of audio these models are unable to distinguish the target signal from the interfering speech.

\vspace{-0.1cm}
\subsection{Ablation study}
\label{sec:ablation}
\vspace{-0.1cm}
\par{\noindent \bf Architecture.} To validate our architectural design, we first compare spectrogram variants and inversion methods in the top half of Table~\ref{tab:ablation}. We experiment with two different types of spectrogram as input for our linear encoder: mel-spectrogram (as in \cite{lavoce}) and linear spectrogram. We find that the mel-spectrogram input yields better results across all metrics. In addition, we compare \emph{C-HiFi-GAN} with two common spectrogram inversion methods -- Griffin-Lim~\cite{DBLP:conf/icassp/GriffinL83}, 
and noisy phase reconstruction (as in \cite{DBLP:conf/interspeech/GabbaySP18}).
We find that both methods are slower than \emph{C-HiFi-GAN} and are unable to match its quality and intelligibility. 
In the bottom half of Table~\ref{tab:ablation}, in an attempt to eliminate the algorithm latency introduced by the mel-spectrogram input (15~ms), we further propose three causal audio encoders - a causal mel-spectrogram that leverages the causal padding trick, and two causal 1D ResNets that operate on the raw audio instead (see Section \ref{sec:method_rt_lavoce}). We conclude that the 100~Hz ResNet achieves the best results and even outperforms the original mel-spectrogram encoder, while reducing the algorithm latency by 15~ms. Finally, we show that increasing the left context length of the Emformer from 32 to 64 yields considerable improvements in performance while having negligible impact on the latency.

\par{\noindent \bf Causal vs. non-causal components.} We compare the causal and non-causal versions of the temporal encoder (Emformer and Transformer) and neural vocoder (\emph{C-HiFi-GAN} and HiFi-GAN) using two inference modes in Table~\ref{tab:on_off_inference}. In offline inference, the enhanced speech is computed using the entire audio-visual input, as in \cite{lavoce}. In online inference, which is the focus of our work, the enhanced audio is generated frame by frame and can only leverage past inputs, as in Figure~\ref{overview_fig}. As expected, we observe that the Transformer and HiFi-GAN surpass their causal counterparts in offline inference, but underperform when generating audio online, since they can no longer leverage future information. In contrast, the Emformer and \emph{C-HiFi-GAN} achieve the same performance in both inference modes since they leverage only past frames, and achieve the best online inference results by a wide margin. This highlights the need to design causal modules, rather than na\"ively leveraging the non-causal components presented in \cite{lavoce} for real-time enhancement. 

\vspace{-0.1cm}
\subsection{End-to-end real-time enhancement}
\vspace{-0.1cm}
\label{sec:real-time}

\par{\noindent \bf Low-latency mouth cropping.} In the experiments presented above, we obtain the mouth crops following the pipeline presented in LA-VocE~\cite{lavoce}, which has an average latency of 92.65\,ms. In Table~\ref{tab:lightweight_crop}, we explore whether our trained models can produce high-quality speech using mouth crops generated by a causal low-latency (7.3\,ms per frame) mouth cropping pipeline based on MediaPipe~\cite{DBLP:journals/corr/abs-1906-08172}. We find that the performance difference between these two methods is marginal (roughly 2\,\%), which shows that RT-LA-VocE can effectively adapt to the videos produced by MediaPipe without a significant loss in performance. 

\par{\noindent \bf Total processing latency.} Given RT-LA-VocE's and MediaPipe's latencies per frame (20.88\,ms and 7.27\,ms, respectively), our end-to-end pipeline has a total processing latency of 28.15\,ms per frame. This is considerably lower than the time it takes to obtain the next frame (\ie, 40\,ms), meaning that we can perform real-time enhancement without accumulating delays at each time-step.
\vspace{-0.3cm}
\section{Conclusion}
In this paper, we present RT-LA-VocE, a real-time audio-visual speech enhancement model that adapts the original LA-VocE~\cite{lavoce} for causal inference. 
Our proposed architecture sets a new state-of-the-art for real-time AVSE, and is competitive with the original LA-VocE during offline inference, despite not relying on future information. 
Furthermore, our findings show that RT-LA-VocE can be leveraged via a server-side GPU to enhance noisy speech in live video streams with minimal processing latency ($<$~30~ms).
\vspace{0.2cm}

\section{References}
\AtNextBibliography{\ninept}
\setlength\bibitemsep{0.65\itemsep}
\printbibliography[heading=none]
\end{document}